\documentclass[natbib]{svjour3}                     
\usepackage{graphicx}

\usepackage{hyperref}
\usepackage{amssymb}
\usepackage{color}
\usepackage{bm}
\AtBeginDocument{}

\journalname{Celest. Mech. Dyn. Astr.}

\begin{document}

\title{Influence of a second satellite on the rotational dynamics of an oblate moon}

\author{Mariusz Tarnopolski}

\institute{M. Tarnopolski \at
              Astronomical Observatory of the Jagiellonian University\\ul. Orla 171, 30-244 Krak\'ow, Poland \\
              \email{mariusz.tarnopolski@uj.edu.pl}
}

\date{Received: date / Accepted: date}

\maketitle

\begin{abstract}
The gravitational influence of a second satellite on the rotation of an oblate moon is numerically examined. A simplified model, assuming the axis of rotation perpendicular to the (Keplerian) orbit plane, is derived. The differences between the two models, i.e. in the absence and presence of the second satellite, are investigated via bifurcation diagrams and by evolving compact sets of initial conditions in the phase space. It turns out that the presence of another satellite causes some trajectories, that were regular in its absence, to become chaotic. Moreover, the highly structured picture revealed by the bifurcation diagrams in dependence on the eccentricity of the oblate body's orbit is destroyed when the gravitational influence is included, and the periodicities and critical curves are destroyed as well. For demonstrative purposes, focus is laid on parameters of the Saturn-Titan-Hyperion system, and on oblate satellites on low-eccentric orbits, i.e. $e\approx 0.005$.
\end{abstract}

\section{Introduction}\label{intro}

Saturn's seventh moon, Hyperion (also known as Saturn VII), was discovered in the XIX century by \citet{bond} and \citet{lassel}, but only due to Voyager~2 \citep{smith} and Cassini \citep{thomas10} missions it became apparent that it is the biggest known highly aspherical celestial body in the Solar System, with a highly elongated shape and dimensions $360\times 266\times 205$ km. Since the rotational state of Hyperion was predicted to remain in the chaotic zone \citep{wisdom} based on the spin-orbit coupling theory \citep{goldreich}, further analyses and observations, regarding Hyperion as well as other Solar System satellites, were conducted on a regular basis.

Hyperion's long-term observations were carried out twice in the post Voyager~2 era. In 1987, \citet{klav,klav2} performed photometric $R$ band observations over a timespan of more than 50 days, resulting in 38 high-quality data points. In 1999 and 2000, \citet{devyatkin} conducted $C$ (integral), $B$, $V$ and $R$ band observations. The objective of both analyses was to determine whether Hyperion's rotation is chaotic and to fit a solution of the equation of motion to the observations. To the best of the author's knowledge \citetext{Melnikov, priv.\ comm.} there were no other long-term observations that resulted in a lightcurve allowing the determination of Hyperion's rotational state (see also \citealt{strugnell} and \citealt{dourneau} for a list of earlier observations). Although, shortly after the Cassini 2005 passage a ground-based $BVR$ photometry was conducted \citep{hicks}, resulting in 6 nights of measurements (and additional 3 nights of $R$ photometry alone) over a month-long period. Unfortunately, this data was greatly undersampled and period fitting procedures yielded several plausible solutions.

The theoretical and numerical treatment of the rotational dynamics of an oblate satellite have been performed widely. After the seminal paper of \citet{wisdom}, \citet{boyd} applied the method of close returns to a sparse and short-term simulated observations of Hyperion's lightcurve. \citet{black} performed numerical experiments using the full set of Euler equations to model long-term dynamical evolution. \citet{beletskii} considered a number of models, including the gravitational, magnetic and tidal moments as well as rotation in gravitational field of two centers. A model with a tidal torque was examined analytically using Melnikov's integrals and assymptotic methods \citep{khan}. The stability of resonances with application to the Solar System satellites was inferred based on a series expansion of the terms in the equation of rotational motion \citep{celletti1,celletti2}. The Lyapunov exponents and spectra were exhaustively examined for a number of satellites\footnote{In particular, Lyapunov times for Hyperion ranged from $1.5\times T$ to $7\times T$, where $T=21.28\,{\rm d}$ is the orbital period.} \citep{shevchenko1,shevchenko,kouprianov1,kouprianov2}. A model of an oblate satellite with dissipation was used to examine the basins of attraction in case of low eccentricities, especially with application to the Moon \citep{celletti3}. The dynamical stability was examined for all known satellites by \citet{melnikov}. Again the dynamical modeling using the full Euler equations was conducted by \citet{harbison}, who also analyzed the moments of inertia in light of the precessional period. Finally, \citet{tarnopolski} argued that in order to extract a maximal Lyapunov exponent from the photometric lightcurve of Hyperion, at least one year of dense data is required.

The orbital dynamics of Hyperion in the Saturn-Titan-Hyperion system (see Table~\ref{tbl1} for some physical parameters) have been exhaustively examined due to the interesting 4:3 mean motion resonance between Hyperion and Titan \citep{peale,taylor2,stellmacher,rein}. While the impact of Titan's gravitation on Hyperion's orbit has been established \citep{taylor} and the stability of the resonance has been considered in great detail \citep{colombo,bevilacqua}, introduction of the gravitational impact of a secondary body on the rotation of an oblate satellite has been done before for nearly spherical bodies such as Venus \citep{beletskii2} or low-eccentric orbits in the Pluto-Charon system \citep{correia}. Herein, numerical integrations will be performed within the chaotic zone of the Saturn-Titan-Hyperion system with parameters $\omega^2$ and $e$ such that the perturbation techniques are not valid \citep{maciejewski}, which to the best of the author's knowlege has not yet been examined and hence is the aim of this work, which is general enough to be applicable to moons other than Hyperion. To focus attention, throughout the analysis the parameters are set to those of the Saturn-Titan-Hyperion system unless otherwise stated, but low-eccentricity and low-oblateness cases are also investigated for comparison.

\begin{table}[h]
\small
\caption{Physical parameters of the Saturn-Titan-Hyperion system.}
\label{tbl1}
\centering
\begin{tabular}{@{}cccc@{}}
\hline
Parameter & Symbol & Value & Reference \\
\hline
Saturn's mass & $M$ & $5.68\cdot 10^{26}\,{\rm kg}$ & \citet{jacobson} \\
Titan's mass & $m_1$ & $1.35\cdot 10^{23}\,{\rm kg}$ & \citet{jacobson} \\
 & $m_1/M$ & $2.4\cdot 10^{-4}$ &  \\
Hyperion's major semi-axis & $a$ & 1 429 600 km & \citet{seidelmann,thomas07} \\
Titan's major semi-axis & $a_0$ & 1 221 865 km & \url{http://ssd.jpl.nasa.gov/?sat_elem} \\
 & $a_0/a$ & 0.855 &  \\
Hyperion's oblateness & $\omega^2$ & 0.79 & \citet{wisdom} \\
Hyperion's eccentricity & $e$   & 0.1 & \citet{wisdom} \\
Hyperion's orbital period & $T$ & 21.3 d & \citet{thomas07} \\
\hline
\end{tabular}
\end{table}

This paper is organized in the following manner. In Sect.~\ref{model} the rotational models in case of the absence and presence of a second satellite's gravitation are derived. In Sect.~\ref{phase} the phase space is briefly described. Section~\ref{meth} presents the methods used: the correlation dimension and its benchmark testing, and the bifurcation diagrams. The results are presented in Sect.~\ref{res}, which is followed by discussion and conclusions gathered in Sect.~\ref{disc}. The computer algebra system \textsc{mathematica}\textsuperscript{\textregistered} is applied throughout this paper.

\section{Models}\label{model}

\subsection{Rotational model of an oblate moon}\label{hyperion}

\begin{figure*}
\includegraphics[width=\textwidth]{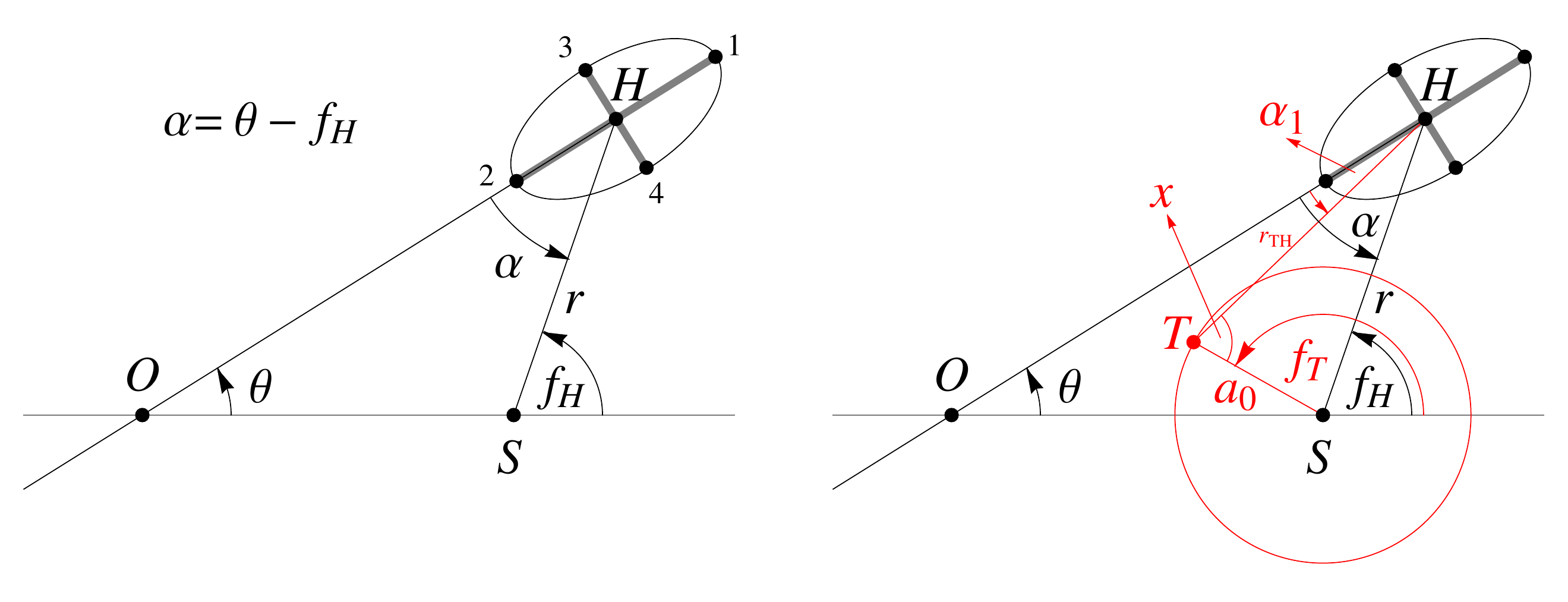}
\caption{{\bf Left:} Rotational model of an oblate moon. {\bf Right:} Geometry of the model including the orbital motion of a second satellite. $S$, $T$ and $H$ stand exemplary for Saturn, Titan and Hyperion (center of mass), respectively. See text for explanation of the remaining symbols.}.
\label{fig1}
\end{figure*}

The equation of rotational motion can be derived based on the following assumptions \citep{greiner}:
\begin{enumerate}
\item the orbit of the satellite around the planet is Keplerian with eccentricity $e$ and major semi-axis $a$:
\begin{equation}
r=\frac{a\left(1-e^2\right)}{1+e\cos f_H},
\label{eq1}
\end{equation}
where $f_H$ is the true anomaly given by
\begin{equation}
\dot{f}_H=\frac{\sqrt{GM}}{\left[a\left(1-e^2\right)\right]^{3/2}}\left(1+e\cos f_H\right)^2,
\label{eq2}
\end{equation}
with $M$ the mass of the planet and the overdot denotes differentation with respect to time;
\item in general, the physical model of the satellite is a triaxial ellipsoid; however, to simplify calculations, the satellite is simulated by a double dumbbell with four mass points 1 to 4 (see Fig.~\ref{fig1}) with equal mass $m$ arranged in the orbital plane. The principal moments of inertia are $A>B>C$;
\item the satellite's spin axis is fixed and perpendicular to the orbit plane; the spin axis is aligned with the shortest physical axis, i.e. the one corresponding to the largest principal moment of inertia.
\end{enumerate}
In case of Hyperion, the first assumption is not precisely valid, as it is well known that due to gravitational interaction with Titan the eccentricity of Hyperion oscillates from $\sim 0.08$ to \mbox{$\sim 0.12$} with an 18.8-year period \citep{taylor}. However, as the analysis herein is performed on a time span much shorter than this period (i.e., $<1\,{\rm yr}$), the effect of this interaction will be negligible and as such is omitted \citep{black,shevchenko}. The second assumption, while might look like an oversimplification at first, does not affect the final equation of motion, which is the same as the one obtained directly from the Euler equations (\citealt{danby}; see also Appendix~\ref{app1} for a remark on the moments of inertia in both models). The third assumption is justified for most satellites as the angular momentum is assumed to be constant with great accuracy. However, it should be noted that \citet{wisdom} showed that the chaotic state is attitude unstable, and also the analysis of Voyager 2 images showed that the axis of rotation was far from being perpendicular to the orbital plane. Therefore, the models derived herein are a first approximation that will be expected to give initial insight into the dynamics of the satellite, and the paramateres corresponding of the Saturn-Titan-Hyperion system are used for demonstrative reasons.

Defining the oblateness as $\omega^2=\frac{3(B-A)}{C}$, and choosing the units so that the orbital period $T$ is equal to $2\pi$ and the major semi-axis $a=1$ (which implies through Kepler's third law $GM=1$, and that the orbital mean motion $n=1$), eventually the equation of motion takes the form
\begin{equation}
\ddot{\theta}+\frac{\omega^2}{2r^3}\sin 2\left(\theta-f_H\right)=0,
\label{eq11}
\end{equation}
where time is dimensionless and $\dot{\theta}$ is measured in units of $n$, with Eq.~(\ref{eq1}) for the orbit in the form
\begin{equation}
r=\frac{\left(1-e^2\right)}{1+e\cos f_H},
\label{eq12}
\end{equation}
and Eq.~(\ref{eq2}) for the true anomaly yields
\begin{equation}
\dot{f}_H=\frac{1}{\left(1-e^2\right)^{3/2}}\left(1+e\cos f_H\right)^2.
\label{eq13}
\end{equation}
Moreover, transforming Eq.~(\ref{eq11}) so that $f_H$ is the independent variable leads to the famous Beletskii equation \citep{beletskii3}, which was shown to be non-integrable \citep{maciejewski}.

\subsection{Introducing a second satellite}\label{titan}

Herein, a second satellite is assumed to revolve around the planet on a circular orbit, with radius $a_0$, in the same plane as the oblate moon. Based on Eq.~(\ref{eq13}), the true anomaly depends linearly on time:
\begin{equation}
f_T=\frac{1}{a_0^{3/2}}t.
\label{eq14}
\end{equation}
From the triangle $TSH$ (see Fig.~{\ref{fig1}}) one obtains that the distance between the two satellites is equal to
\begin{equation}
r_{TH}=r\sqrt{1-\frac{2a_0}{r}\cos\left(f_T-f_H\right)+\left(\frac{a_0}{r}\right)^2}.
\label{eq15}
\end{equation}
The angle $\alpha_1$ is also required. Using again the triangle $TSH$ one finds
\begin{equation}
\alpha_1=x+f_T-\theta-\pi.
\label{eq16}
\end{equation}
The angle $x$ can be found by applying the law of cosines to the same triangle $TSH$, what gives
\begin{equation}
x=\arccos\left(\frac{r_{TH}^2+a_0^2-r^2}{2r_{TH}a_0}\right).
\label{eq17}
\end{equation}
Inserting Eq.~(\ref{eq17}) into Eq.~(\ref{eq16}) one arrives at the formula for $\alpha_1$.

Finally (see Appendix~\ref{app2}), one obtains the following equation of motion:
\begin{equation}
\begin{array}{l}
\ddot{\theta}=\frac{\omega^2}{2}\Big\{\frac{\sin 2\left(f_H-\theta\right)}{r^3}-\\
\textcolor{white}{\ddot{\theta}=} \frac{m_1/M}{r_{TH}^3}\sin 2\left[\arccos\left(\frac{a_0-r\cos\left(f_T-f_H\right)}{r_{TH}}\right)+f_T-\theta\right]\Big\}.
\end{array}
\label{eq21}
\end{equation}
The initial conditions (ICs) for the true anomalies will be assumed throughout to be $f_H(0)=f_T(0)=0$. The backward differentiation formula (BDF) is employed for numerical integrations \citep{ascher}.

\section{Phase space properties}\label{phase}

In this Section the structure of the phase space of the dynamical system given by Eq.~(\ref{eq11})--(\ref{eq13}) is briefly described. This will allow an insight into how does the gravitational interaction with the second satellite influence the oblate moon's rotation.

\begin{figure}[h]
\begin{center}
\includegraphics[width=\columnwidth]{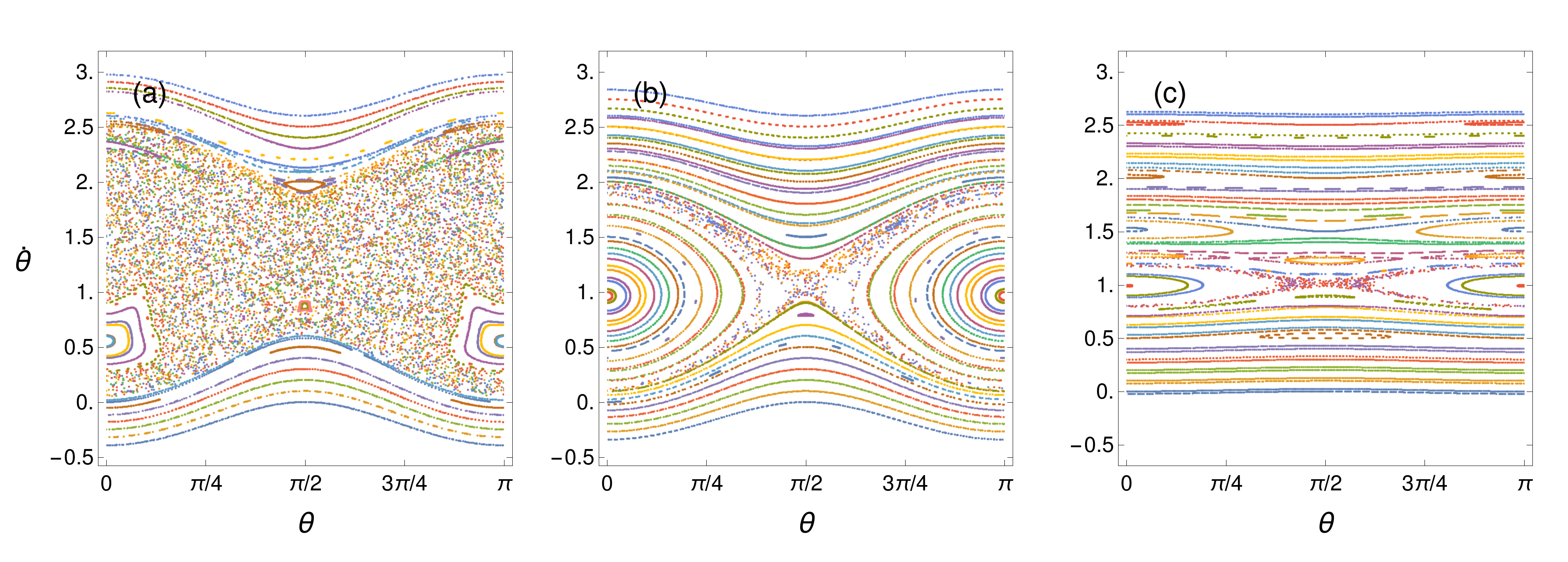}
\end{center}
\caption{Phase space in stroboscopic variables for (a) $e=0.1,\,\omega^2=0.79$, (b) $e=0.005,\,\omega^2=0.79$, (c) $e=0.1,\,\omega^2=0.04$. Different colours correspond to different trajectories (ICs). Eq.~(\ref{eq11})---(\ref{eq13}) were integrated for 2,000 dimensionless time units, resulting in 319 points for each trajectory. The initial conditions were $(0,\dot{\theta}_0)$ and $(\pi/2,\dot{\theta}_0)$, where $\dot{\theta}_0\in\{0.0,0.1,\ldots,2.5,2.6\}$.}
\label{phaseplot}
\end{figure}

The phase space is 3-dimensional: $\Omega=\{(\theta,\dot{\theta},f_H):\theta\in\mathbb{R}\,{\rm mod}\,2\pi,\dot{\theta}\in\mathbb{R},f_H\in\mathbb{R}\,{\rm mod}\,2\pi\}$. But $f_H$ is a regular, $2\pi$-periodic function and does not carry much information. Moreover, in dimensionless units the orbital period of the oblate moon is also equal to $2\pi$. Hence, a Poincar\'e surface of section, constructed by taking the values of $(\theta,\dot{\theta})$ with a time step of $2\pi$, i.e. employing stroboscopic variables, provides insight into the rotational dynamics. Furthermore, the rotation of the satellite by $180^\circ$ (i.e., $\theta\rightarrow\theta+\pi$) gives an equivalent configuration, hence $\theta$ can be confined to the interval $[0,\pi)$. Such surfaces of section are shown in Fig.~\ref{phaseplot}.

As is common in Hamiltonian systems, the phase space is divided into regions occupied with chaotic trajectories, and regions of regular (periodic or quasiperiodic) motion. There are also motions called sticky orbits, when the trajectory initially behaves in a regular manner and diverges into the chaotic zone after some time. (See also \citealt{melnikov14} for the emergence of strange attractors when dissipation is introduced.) The phase space in Fig.~\ref{phaseplot}(a) is dominated by a large chaotic sea, formed by merging the chaotic zones surrounding spin-orbit resonances from $p$=1:2 to $p$=2:1 when $\omega^2$ increases \citep{wisdom}. Quasiperiodic motions are represented by smooth curves and by closed curves, e.g. the ones surrounding the synchronous $p$=1:1 resonance (see Table~1 in \citealt{black} for locations of the surviving resonances). When the IC is located near the boundary between regular motion and the chaotic sea, sticky motion occurs. A narrow chaotic zone is present also in Fig.~\ref{phaseplot} (b) and (c) obtained for smaller values of the eccentricity and oblateness, respectively. In fact, every Solar System satellite has a chaotic period in its past \citep{ency}.

\section{Methods}\label{meth}

\subsection{Correlation dimension}\label{corr}

The algorithm and programme for computing the correlation dimension are briefly described in Sect.~\ref{alg}, which is followed by the description of methodology and discussion of the results of the benchmark testing in Sect.~\ref{benchmark}.

\subsubsection{Algorithm}\label{alg}

A fractal dimension (or, more precisely, the Hausdorff dimension; \citealt{hausdorff,theiler,ott}) is often measured with the correlation dimension, $d_C$ \citep{grassberger,grassberger2,theiler,alligood,ott}, which takes into account the local densities of the points in the examined dataset. For usual 1D, 2D or 3D cases the $d_C$ is equal to 1, 2 and 3, respectively. Typically, a fractional correlation dimension is obtained for fractals \citep{mandelbrot}.

The correlation dimension is defined as
\begin{equation}
d_C=\lim_{R\rightarrow 0}\frac{\ln C(R)}{\ln R},
\label{eqA}
\end{equation}
with the estimate for the correlation function $C(R)$ as
\begin{equation}
C(R)=\frac{1}{N^2}\sum_{i=1}^N\sum_{j=i+1}^N \Theta\left(R-||x_i-x_j||\right),
\label{eqB}
\end{equation}
where the Heaviside step function $\Theta$ adds to $C(R)$ only points $x_i$ in a distance smaller than $R$ from $x_j$ and vice versa. The total number of points is denoted by $N$, and the usual Euclidean distance, $||{\bm .}||$, is employed. The limit in Eq.~(\ref{eqA}) is attained by using multiple values of $R$ and fitting a straight line to the linear part of the obtained dependency. The correlation dimension is estimated as the slope of the linear regression. The computations in this work were performed using the \textsc{python} code from \citep{tarnopolski1}, with a slight modification so that $\ln R$, instead of $R$, is uniformly sampled with a step $\Delta\ln R$. Throughout this paper, when the $d_C$ is considered, $N$ is set to be 10,201 (see Sect.~\ref{res2}).

\subsubsection{Benchmark testing}\label{benchmark}

In order to verify the reliability of the correlation dimension algorithm, benchmark testing is performed on 128 realisations of uniform sampling with $N$ points in each of the regions I and II defined as follows: region I is a unit square, and region II is a unit square without a circle of radius $1/4$ placed at the center of the square. Next, the $d_C$ is computed as described in Sect.~\ref{alg}, with $\ln R\in[-8,-2]$ and $\Delta \ln R=0.5$. The results are gathered in Table~\ref{tbl2}, from which it follows that the estimated $d_C$ is very close to the correct dimension expected for an Euclidean 2-dimensional space.
\begin{table}[h]
\small
\caption{Results of the correlation dimension benchmark testing for uniformly sampled regions I and II; $\sigma$ denotes the standard deviation of the sample.}
\label{tbl2}
\centering
\begin{tabular}{@{}ccc@{}}
\hline
Region & $\langle d_C\rangle$ & $\sigma$ \\
\hline
I   & 1.988 & 0.027 \\
II  & 1.988 & 0.023 \\
\hline
\end{tabular}
\end{table}

To validate the performance when clustering is introduced, regions I and II are uniformly sampled with 9,000 points, and the remaining 1,201 points are introduced in a circular region with radius equal to 1/8 (which is overlaid with points in the unit square), randomly chosen at each of the 128 realisations and lying entirely within region I or II. The results are gathered in Table~\ref{tbl3}. The correlation dimensions are close to the expected value of 2, but slightly smaller than they were when there was no clustering. 
\begin{table}[h]
\small
\caption{Results of the correlation dimension benchmark testing when clustering is introduced.}
\label{tbl3}
\centering
\begin{tabular}{@{}ccc@{}}
\hline
Region & $\langle d_C\rangle$ & $\sigma$ \\
\hline
I   & 1.981 & 0.026 \\
II  & 1.974 & 0.022 \\
\hline
\end{tabular}
\end{table}

\begin{figure}[h]
\includegraphics[width=\columnwidth]{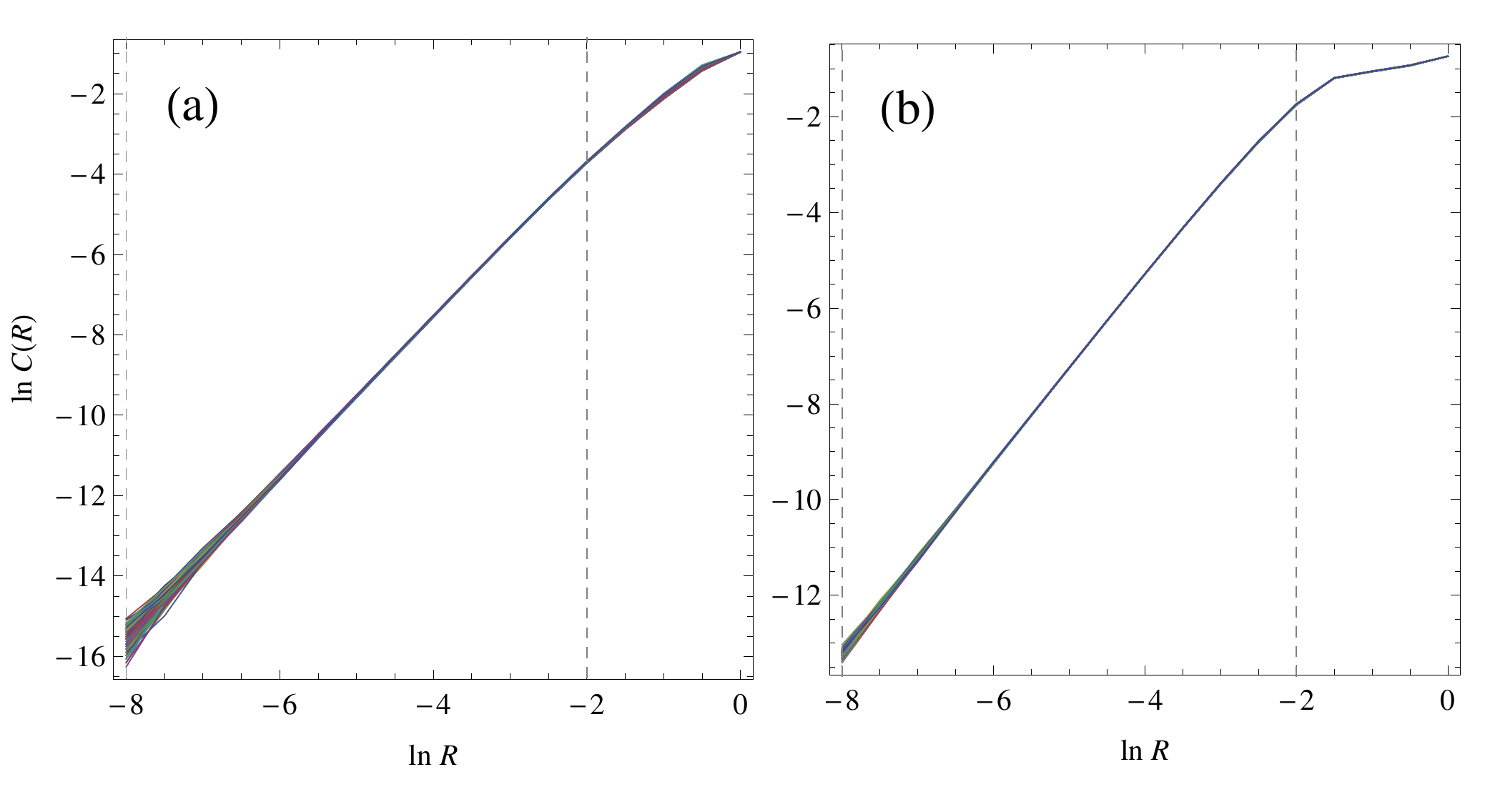}
\caption{Estimation of the $d_C$ for 128 realisations of (a) a uniformly sampled unit square and (b) for 2,101 points in a unit square overlaid with 8,000 points in a $0.25\times 0.25$ square (clustering). The vertical dashed lines mark the region where linear regression was performed. Note different scales on the vertical axis. When clustering is introduced the $\ln C(R)$ vs. $\ln R$ dependency is broken at some point.}
\label{fig8}
\end{figure}

Finally, 128 realisations of uniform distributions of 2,101 points within a unit square, which was overlaid with 8,000 points distributed uniformly in a $0.25\times 0.25$ square located in one corner of the unit one, were generated. The resulting mean $d_C$ was 1.937, and $\sigma=0.008$. Hence it was shown that clustering might result in a correlation dimension systematically lower than the expected one. Fig.~\ref{fig8} shows the $\ln C(R)$ vs. $\ln R$ relations for a uniformly sampled unit square and for the last numerical experiment on clustering.

\subsection{Bifurcation diagrams}\label{bif}

In dynamical systems theory, a bifurcation occurs when an infinitesimal change of a (nonlinear) parameter governing the system leads to a topological change in its behaviour. Generally speaking, at a bifurcation, the stability of equilibria, periodic orbits or other invariant sets is changed. The theory of bifurcation is a vast field \citep{crawford,lichtenberg,baker,alligood,ott,kuznetsov,peitgen}. Herein focus will be laid on the pitchfork bifurcations, that are present e.g. in the logistic map \citep{may,feigenbaum} and that constitute one of the routes to chaos. Let us consider a system of differential equations in the form $\dot{x}=f(x;\alpha)$, where $\alpha$ is a parameter, and assume that given $x_0$ as an IC, for $\alpha<\alpha_1$ the orbit is 1-periodic. A bifurcation at $\alpha=\alpha_1$ is a point where the trajectory begins to be 2-periodic and maintains its periodicity up to $\alpha=\alpha_2$. Similarly, at $\alpha=\alpha_2$ a bifurcation occurs on each of the two branches, hence the orbit becomes 4-periodic. This scheme, called a period-doubling (pitchfork) bifurcation cascade, continues until at $\alpha=\alpha_{\infty}<\infty$ the orbit becomes chaotic. However, in the chaotic zone, $\alpha>\alpha_{\infty}$, windows of periodic motion with arbitrary period occur. E.g., when a 3-periodic trajectory emerges from the chaotic zone it also undergoes the period-doubling, hence produces orbits that are 6-periodic, 12-periodic, and so on. A bifurcation diagram is a diagram illustrating this complex mechanism with respect to the nonlinear parameter $\alpha$. Finally, bifurcations may also occur when $\alpha$ is decreasing (period-halving bifurcations) as well as when $|\alpha|$ is decreasing or increasing.

Usually, on the bifurcation diagrams there appear to be some curves running through the plot in the chaotic region. These are the so called critical curves \citep{peitgen} defined by $x=f^n(x_0;\alpha)$.

\section{Results}\label{res}

\subsection{Correlation dimension}\label{res1}

In this section the influence of the second satellite's absence or presence on the phase space flow is examined. In order to do so, two sets of ICs $(\theta_0,\dot{\theta}_0)$ are chosen:
\begin{enumerate}
\item IC1 -- a total of $101\times 101=10,201$ ICs distributed uniformly (with a step of $10^{-3}$) on a $0.1\times 0.1$ square centered on $(\pi/2,0.55)$;
\item IC2 -- similar to IC1, but centered on $(\pi/2,1.5)$.
\end{enumerate}
IC1 was chosen so that it is located on the edge of the chaotic sea and the domain of quasiperiodic motion [according to \citealt{black}, the 1:2 resonance is located at $(\pi/2,0.9)$; compare also with Fig.~\ref{phaseplot}(a)], and IC2 was chosen so that it lies entirely within the chaotic zone (according to \citet{wisdom}, the 3:2 resonance does not exist).

\begin{figure*}
\includegraphics[width=\textwidth]{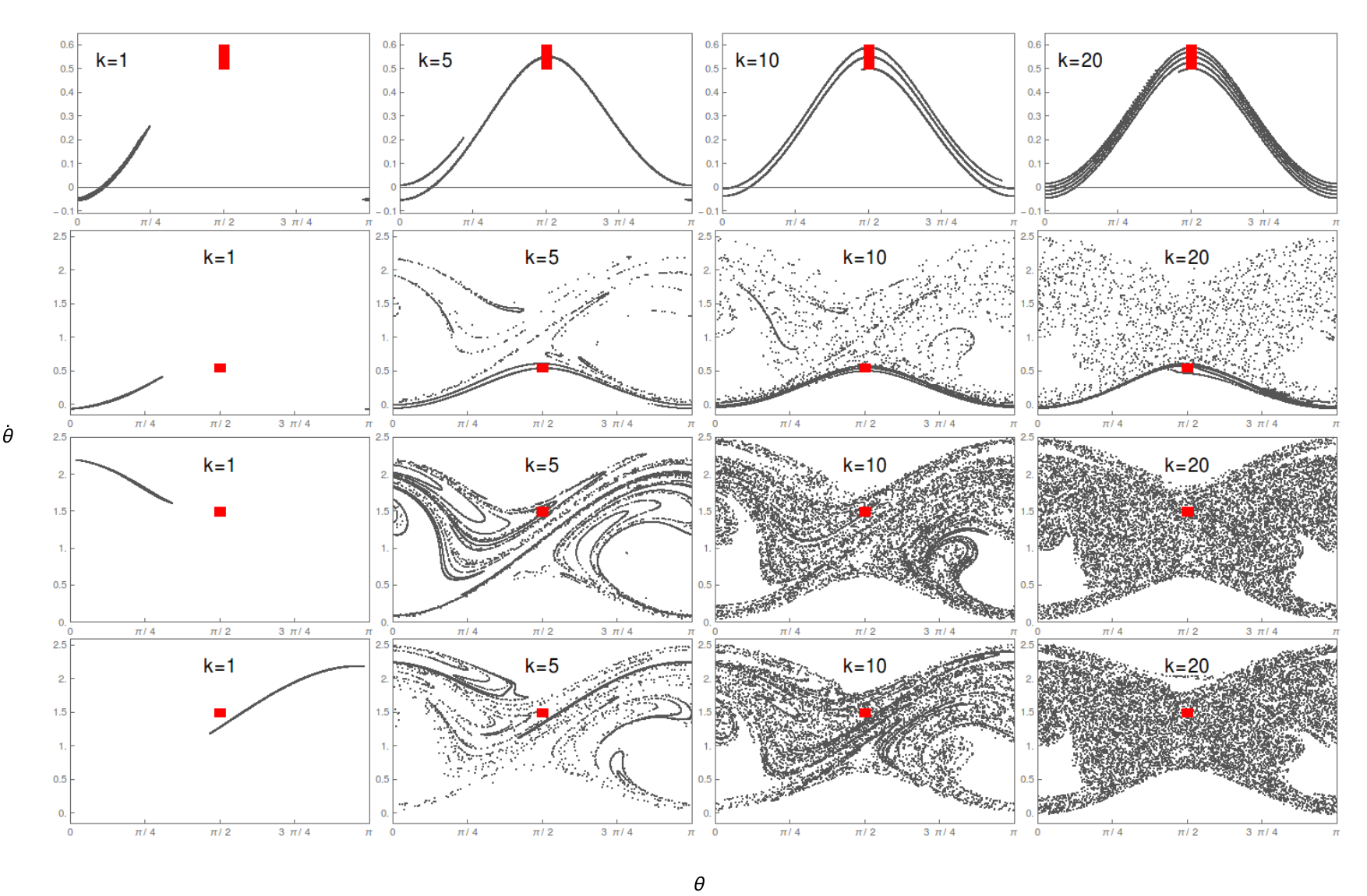}
\caption{The phase space evolution after $k$ revolutions of (from top to bottom): IC1 in case of the second satellite's absence, IC1 but including additional gravitational influence, IC2 in the absence of the second satellite, and IC2 in its presence. The sets IC1 and IC2 are depicted in red in each panel.}
\label{serieplot}
\end{figure*}

The equations of motion (\ref{eq11}) and (\ref{eq21}) are solved numerically for every IC in the sets IC1 and IC2. The integration time is equal to 20 orbital periods of the oblate satellite, i.e. $t_{\rm max}=20\times 2\pi$. After each revolution, starting from $t=0$ (corresponding to the sets IC1 and IC2), the value of $\theta$ and $\dot{\theta}$ is recorded, and the sets to which IC1 and IC2 evolved after $k$ orbital periods are displayed in Fig.~\ref{serieplot}.

\begin{figure}[h]
\includegraphics[width=\columnwidth]{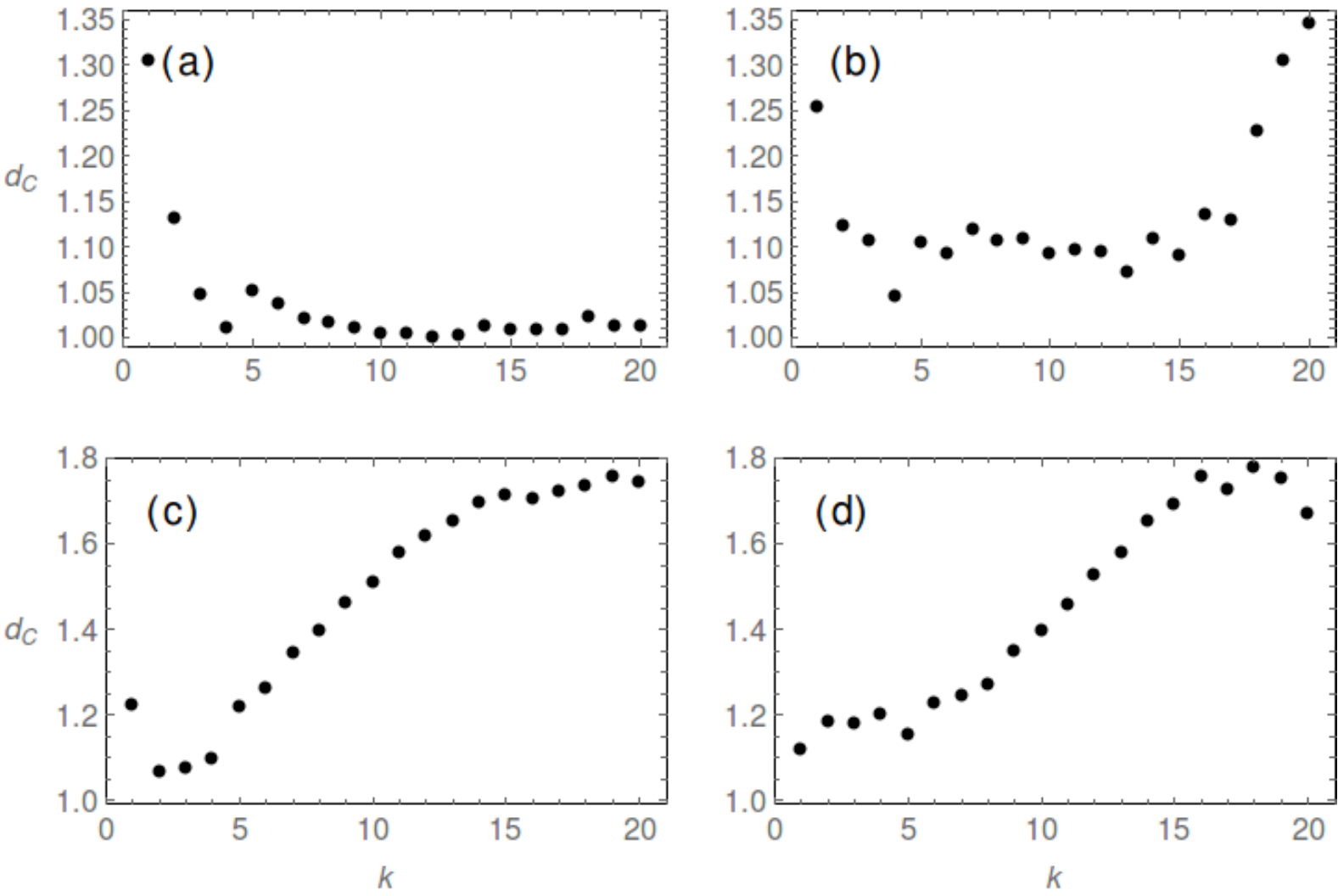}
\caption{The correlation dimension for the sets in Fig.~\ref{serieplot}. Note different scales on the vertical axes on (a)--(b) and (c)--(d).}
\label{fig7}
\end{figure}

Next, the correlation dimensions of these sets are computed according to Sect.~\ref{corr}. The linear regression was performed for $\ln C(R)$ vs. $\ln R$ relation in the region $\ln R\in(-7,\ln R_{\rm th})$, where $R_{\rm th}$ was chosen in each case so that the fit was done only in the linear part of the plot. The results, in graphical form, are displayed in Fig.~\ref{fig7}. Because all orbits corresponding to IC1 and the absence of the second satellite are quasiperiodic, the $d_C$, as expected, plateaus to a value near 1 [Fig.~\ref{fig7}(a)]. However, when its influence is taken into account, the $d_C$ initially tends to a slightly higher value, approximately 1.1, but then starts to rise suddenly [Fig.~\ref{fig7}(b)]. As can be seen in the second row of Fig.~\ref{serieplot}, some of the orbits appear to remain quasiperiodic when the other satellite's gravitation is switched on, but some become chaotic. On the other hand, based on the behaviour of $d_C$ in Fig.~\ref{fig7}(b), one might suspect that sticky chaos is also encountered. While the analysis of different types of rotation---periodic, quasiperiodic, and chaotic, including sticky chaos---is beyond the scope of this work, it is remarkable to note a clear impact of a second satellite on the rotation of an oblate moon: some ICs that would lead to quasiperiodic motion become chaotic.

The evolution of IC2 is very similar in case of both the absence and presence of the second satellite's gravity (third and fourth rows in Fig.~\ref{serieplot}). Also, the $d_C$ behaves in the same manner for both models, reaching a plateau of $d_C\approx 1.75$ after about 16 orbital periods, as shown in Fig.~\ref{fig7}(c) and (d). It is important to note that the value 1.75 cannot be a reliable estimate of the supposed fractality of the attained structure due to the following reasons:
\begin{itemize}
\item both models given by Eq.~(\ref{eq11}) and (\ref{eq21}) are Hamiltonian and hence cannot posses a strange attractor \citep{greiner} that could be characterized by a fractional correlation dimension; assymptotically any set of ICs leading to chaotic orbits should occupy a 2-dimensional subset in the phase space;
\item the number of points, $N=10,201$, used to estimated the $d_C$ is relatively small and hence might bias the outcome \citep{tarnopolski1};
\item local densities exceeding the average density (clustering) affect the correlation dimension such that it is lower than the dimension of the embedding space [see Sect.~\ref{benchmark} and \citep{tarnopolski2}].
\end{itemize}

\subsection{Bifurcation diagrams}\label{res2}

\begin{figure*}[h]
\includegraphics[width=\textwidth]{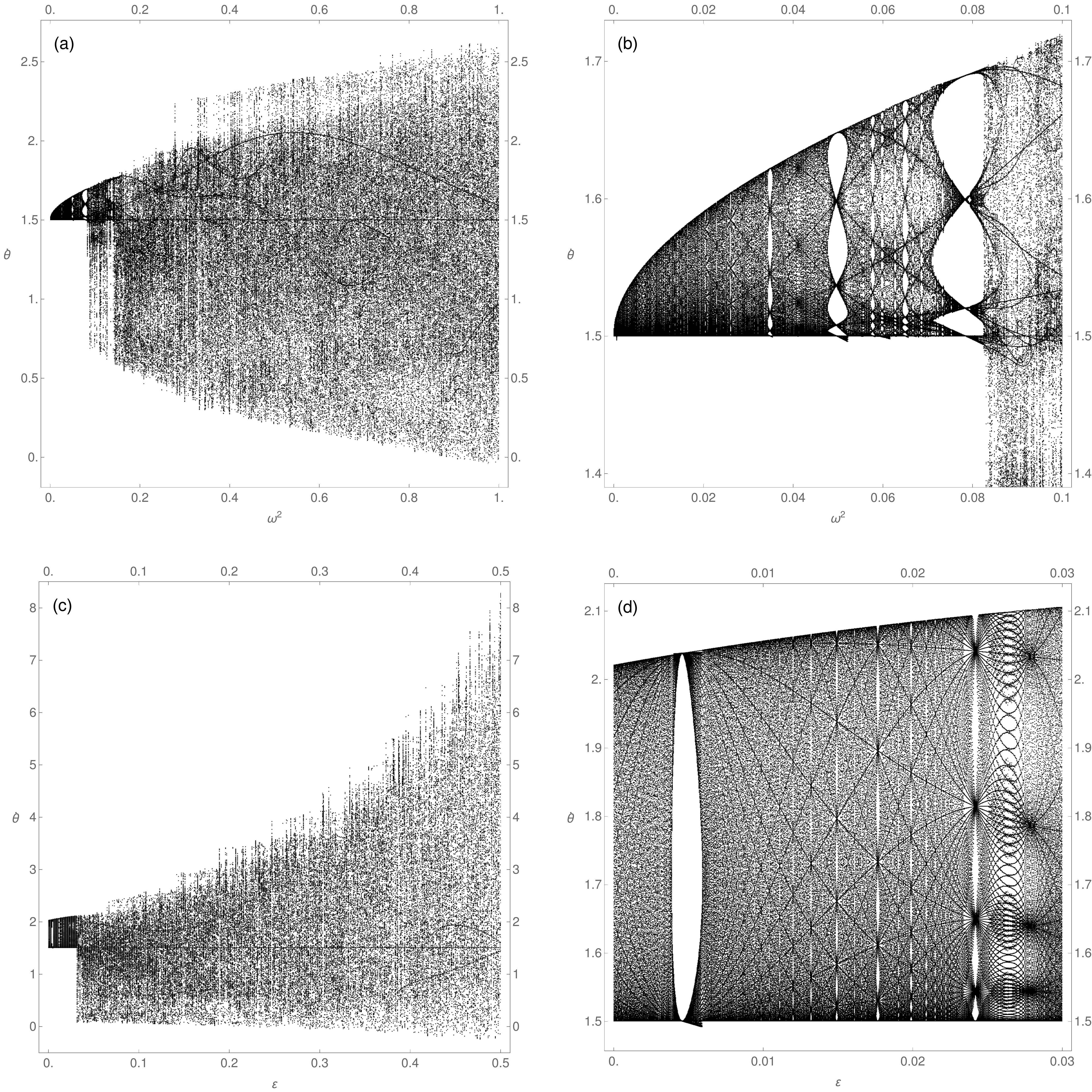}
\caption{Bifurcation diagrams for the case of the absence of the second satellite. In (a)--(b) the eccentricity $e=0.1$, and $\Delta\omega^2=0.001$ and 0.0001, respectively. In (c)--(d) the oblateness $\omega^2=0.79$, and $\Delta e=0.001$ and 0.00002, respectively.}
\label{out1}
\end{figure*}

First, Eq.~(\ref{eq11}), describing the rotation in the absence of a second satellite, is integrated using the initial conditions $(\theta_0,\dot{\theta}_0)=(\pi/2,3/2)$ in the time range $t\in(0,10^4)$, and the values of $\dot{\theta}$ are recorded every revolution, i.e. with a time step of $2\pi$. To obtain the bifurcation diagrams in dependence on the oblateness, the eccentricity $e$ was set to 0.1 and $\omega^2$ was varied. Next, the same procedure was undertaken to obtain the bifurcation diagrams in dependence on the eccentricity, i.e. the oblateness was set to $\omega^2=0.79$ and $e$ was varied. The results are presented in Fig.~\ref{out1}(a) and (c). For relatively small values of $\omega^2$ and $e$, the magnifications in Fig.~\ref{out1}(b) and (d) show a very complex structure with alternating quasiperiodic and periodic orbits with a wide range of periods, e.g. a 4-period at $\omega^2\approx 0.08$ or a 2-period at $e\approx 0.005$ in Figs.~\ref{out1}(b) and (d), respectively. Note that global chaos occurs at $\omega^2\approx 0.083$, which corresponds to $\omega\approx 0.29$, what is in good agreement with the critical value obtained by \citet{wisdom} using the resonance overlap criterion \citep{chirikov,lichtenberg}, $\omega^{\rm RO}=1/\left(2+\sqrt{14e}\right)\approx 0.31$. Hence the bifurcation diagrams confirm the applicability of this criterion to the rotation of an oblate moon.

\begin{figure*}[h]
\includegraphics[width=\textwidth]{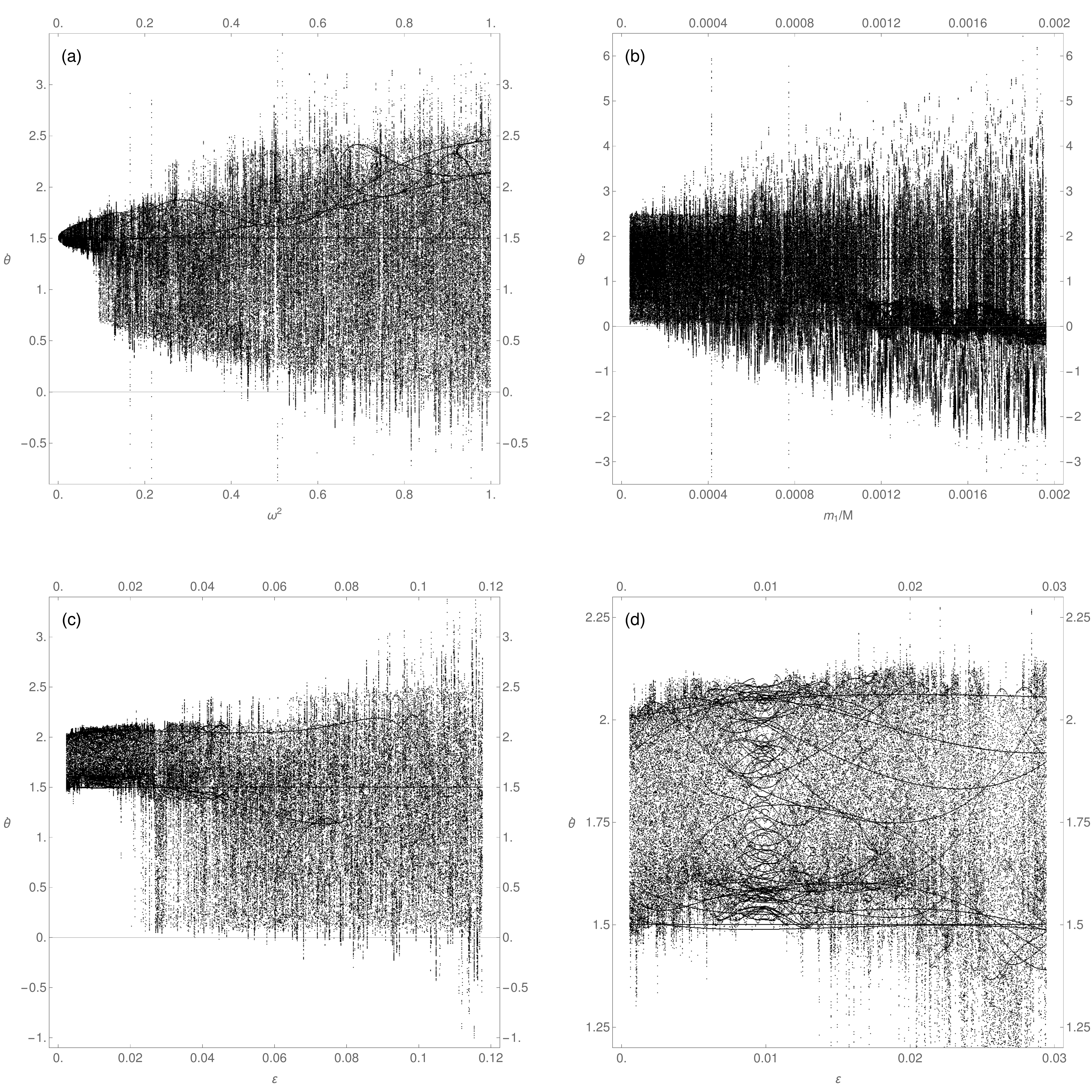}
\caption{Bifurcation diagrams when the gravitational influence of the second satellite is included. In (a) the eccentricity $e=0.1$, the mass ratio $m_1/M=0.00024$ and $\Delta\omega^2=0.001$. In (b) $e=0.1$ and $\omega^2=0.79$ with $\Delta\left(m_1/M\right)=10^{-6}$. In (c)--(d) the oblateness $\omega^2=0.79$, $m_1/M=0.00024$, and $\Delta e=0.0002$ and 0.00005, respectively. Note a change of scale on the horizontal axis in (c) compared to Fig.~\ref{out1}(c).}
\label{out2}
\end{figure*}

In the same way the bifurcation diagrams were obtained using Eq.~(\ref{eq21}) including the gravitational influence of a second satellite. In this case the motion is governed by a third parameter, $m_1/M$, in addition to $\omega^2$ and $e$. The diagram in Fig.~\ref{out2}(a) (where $m_1/M$ was set to 0.00024) is qualitatively similar to the corresponding one in Fig.~\ref{out1}(a), and the dependence on the ratio $m_1/M$ in Fig.~\ref{out2}(b) is mainly structureless. However, there is a significant difference when the oblateness and mass ratio are set to the values from Table~\ref{tbl1}, and the eccentricity $e$ is varied. Note that the range of $e$ in Fig.~\ref{out2}(c) is about four times smaller than in the corresponding Fig.~\ref{out1}(c), as the computational complexity of the problem was much higher when Eq.~(\ref{eq21}) was applied. However, using a more sparse grid it was confirmed that the range of $\dot{\theta}$ also increases with the eccentricity (not shown). A remarkable difference becomes apparent in Fig.~\ref{out2}(d): the image drawn by the bifurcation diagram is less structured compared to Fig.~\ref{out1}(d), and the critical curves are much more tangled. Hence, the impact of a second satellite on the rotation is that the additional body destroys a number of periodic orbits that could occur for low eccentricity values, but for the parameters (i.e., oblateness and eccentricity) of the Saturn-Titan-Hyperion system, its dynamics should also be expected to remain in a chaotic state.

\begin{figure*}[h]
\includegraphics[width=\textwidth]{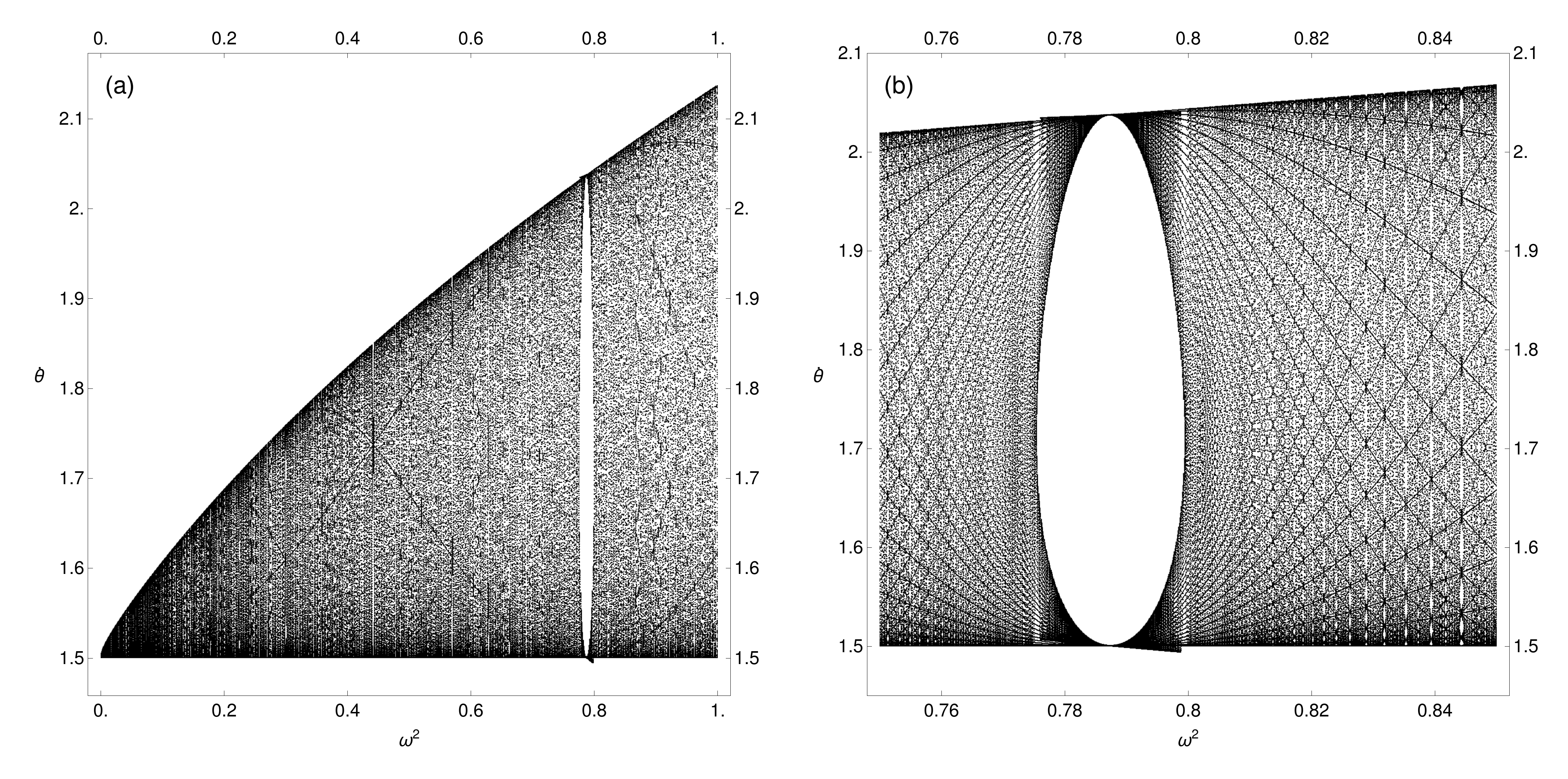}
\caption{Bifurcation diagrams for the case of the absence of the second satellite. The eccentricity $e=0.005$ and (a) $\Delta\omega^2=0.001$; (b) shows the magnification around the 2-periodic window at $\omega^2\approx 0.79$, with $\Delta\omega^2=0.0001$.}
\label{out3}
\end{figure*}

\begin{figure*}[h]
\includegraphics[width=\textwidth]{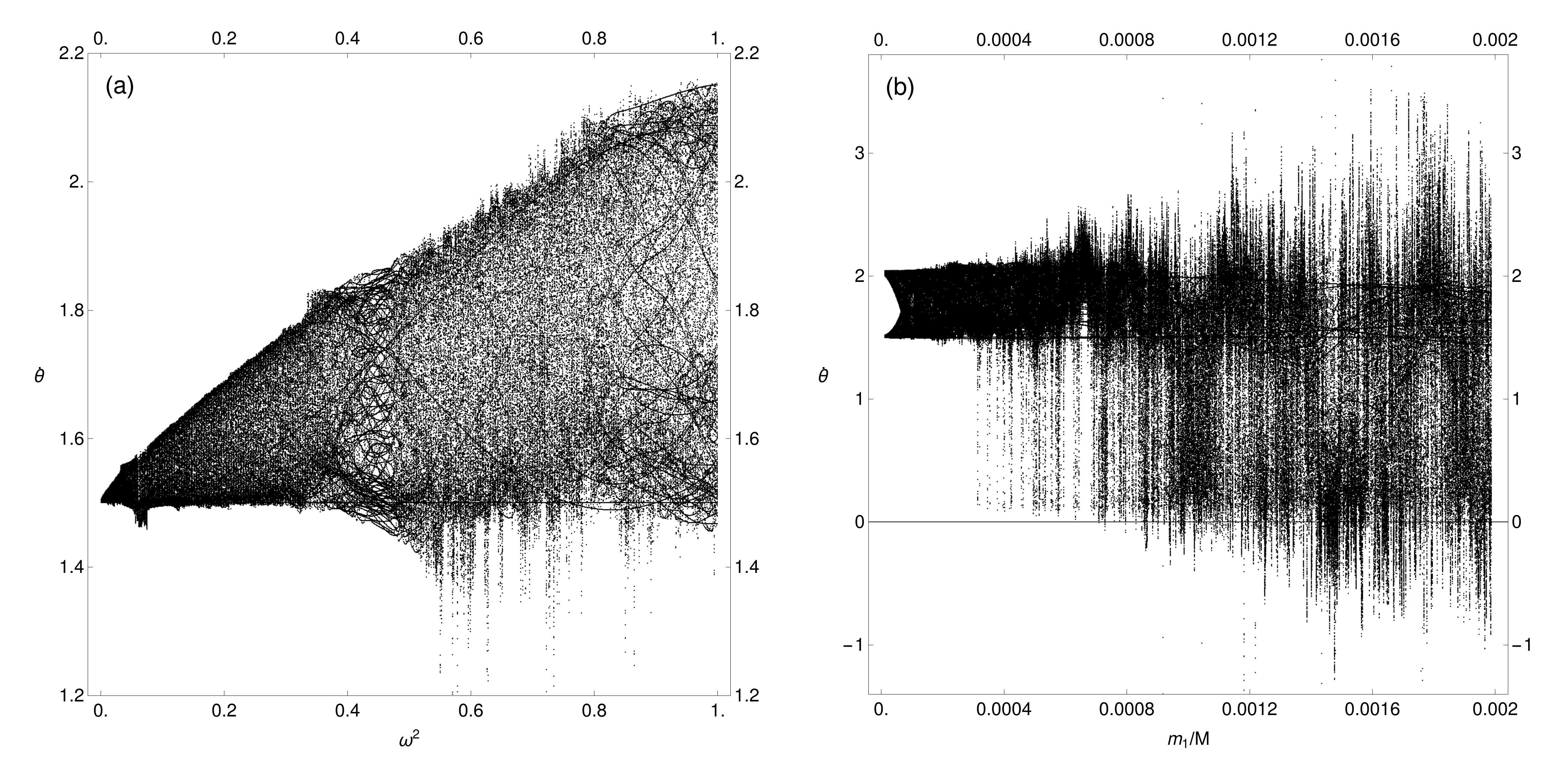}
\caption{Bifurcation diagrams when the gravitational influence of the second satellite is included. In (a) the eccentricity $e=0.005$, the mass ratio $m_1/M=0.00024$ and $\Delta\omega^2=0.001$. In (b) $e=0.005$, $\omega^2=0.79$ and $\Delta\left(m_1/M\right)=10^{-6}$.}
\label{out4}
\end{figure*}

As was pointed out in Sect.~\ref{hyperion}, the model of planar rotation is not applicable to the real Saturn-Titan-Hyperion system as it is attitude unstable. Hence, the bifurcation diagrams are also computed in dependence of $\omega^2$ and the ratio $m_1/M$ with $e=0.005$, which is the mean eccentricity of all Solar System satellites\footnote{\url{http://ssd.jpl.nasa.gov/?sat\_elem}}. Fig.~\ref{out3} shows that for a range of $\omega^2$ the rotation is highly structured in the absence of the second satellite. Interestingly, a distinct 2-periodic window emerges at $\omega^2\approx 0.79$, which surprisingly coincides with the oblateness of Hyperion. Fig.~\ref{out4} shows that the picture becomes much more complex also for such small eccentricity when the gravitational influence of a second satellite is introduced. In particular, the 2-periodic window from Fig.~\ref{out3} has disappeared altogether. Finally, when $\omega^2$ was set to 0.04, the bifurcation diagrams revealed a highly regular structure (not shown).

\section{Discussion and conclusions}\label{disc}

The aim of this paper was to investigate how does the gravitational interaction with a second satellite influence the rotational dynamics of an oblate moon. A simplified model was designed, resulting in the equation on motion given in Eq.~(\ref{eq21}), being basically a perturbation of the well known Eq.~(\ref{eq11}). The derived equation of motion introduces a third parameter, the mass ratio $m_1/M$, additional to te oblateness $\omega^2$ and eccentricity $e$. To allow comparison, two sets of ICs distributed uniformly in a $0.1\times 0.1$ square in the phase space were evolved, in case of the absence and presence of the additional source of gravitation. In case of the set IC1 [centered at $(\pi/2,0.55)$] the difference between the two models is qualitative in nature: when the second satellite was absent all trajectories were quasiperiodic (first row in Fig.~\ref{serieplot}), as indicated by the $d_C=1$ in Fig.~\ref{fig7}(a). Interestingly, when its presence was taken into account, one could observe leaking of the orbits into the chaotic sea (second row in Fig.~\ref{serieplot}). This phenomenon manifests itself also through a higher $d_C$ attained in Fig.~\ref{fig7}(b). Hence, it turns out that an additional satellite has the ability to change quasiperiodic orbits into chaotic ones, i.e. it enlarges the chaotic domain. On the other hand, when the set IC2, located in the center of the chaotic region [an $0.1\times 0.1$ square centered at $(\pi/2,1.5)$], was considered, no long term (assymptotic) differences could be observed (third and fourth rows in Fig.~\ref{serieplot}), and the correlation dimension for both models reached a plateau at $d_C\approx 1.75<2$ [Fig.~\ref{fig7}(c) and (d)], likely due to clustering. However, this is not that surprising, given that the gravitational influence under investigation was three orders of magnitude smaller than the planet's, and that the rotational model in absence of the second satellite is dominated by the chaotic zone, hence it would be highly unexpected for it to have the ability to change chaotic motion into a regular one.

The bifurcation diagrams, especially interesting when $m_1/M$ and $\omega^2$ were fixed and $e$ was varied, when small values of the eccentricity (i.e., $e<0.03$) are considered lead to a conclusion that the regular and highly structured picture [Fig.~\ref{out1}(d)] becomes much more messy, and the transition to chaos occurs for smaller eccentricities than in the case when additional gravitation source is neglected [Fig.~\ref{out2}(d)]. This is consistent with the results of the other method (i.e., evolving the sets IC1 and IC2) in the sense that the second satellite changes regular motion into chaotic. The differences in case when $\omega^2$ was varied was not that much remarkable [Fig.~\ref{out1}(a) and \ref{out2}(a)], and both models lead to chaotic motion when larger $e$ are considered [Fig.~\ref{out1}(c) and \ref{out2}(c)]. The bifurcation diagram in dependence on the ratio $m_1/M$ was mostly structureless [Fig.~\ref{out2}(b)]. Eventually, the destruction of regular rotation caused by the second satellite might be ascribed to the destruction of the invariant tori (\citealt{tabor}; see also \citealt{celletti1} and references therein).

Finally, when $e$ was set to the mean eccentricity of all Solar System satellites (i.e., $e=0.005$), the highly structured bifurcation diagram displayed in Fig.~\ref{out3} also got destroyed and became much more tangled, as shown in Fig.~\ref{out4}. To conclude, the derived simplified model of a second satellite's influence on rotational dynamics of an oblate satellite implies that:
\begin{enumerate}
\item the additional source of gravitation can change some regular orbits into chaotic ones, and
\item destroys the regularity, particularly the periodicities and critical curves, in the bifurcation diagram for small eccentricities $e<0.03$.
\end{enumerate}

\begin{acknowledgements}
The author acknowledges support in form of a special scholarship of Marian Smoluchowski Scientific Consortium Matter-Energy-Future from KNOW funding, grant number KNOW/48/SS/PC/2015.
\end{acknowledgements}

\appendix

	\section{Moments of inertia}\label{app1}

Assuming the oblate moon is a triaxial ellipsoid with axes $a>b>c$ and mass $M=\frac{4\pi}{3}\rho abc$, the respective principal moments of inertia $A<B<C$ are
\begin{equation}
A=\frac{M}{5}(b^2+c^2),\quad B=\frac{M}{5}(a^2+c^2), \quad C=\frac{M}{5}(a^2+b^2).
\label{eqA1}
\end{equation}
In the double dumbbell model (here $c=0$):
\begin{equation}
A=\frac{m}{2}b^2,\quad B=\frac{m}{2}a^2, \quad C=\frac{m}{2}(a^2+b^2).
\label{eqA2}
\end{equation}
Comparing Eq.~(\ref{eqA1}) and (\ref{eqA2}), the masses should obey $m=\frac{2}{5}M$ for the moments of inertia for the ellipsoid model to contain the double dumbbell model. However, this does not affect the analysis performed here because the equations of motion (\ref{eq11}) and (\ref{eq21}) depend only on a dimensionless parameter $\omega^2=\frac{3(B-A)}{C}$ which takes the same value for both models (with $4m=M$).

	\section{Equations of motion}\label{app2}

\subsection{Rotational model of an oblate moon}

Denote by ${\bf e}_1$ the body-fixed axis joining points 2 and 1, and by ${\bf e}_2$ the axis joining points 4 and 3 (see Fig.~\ref{fig1}). Thence, ${\bf e}_3={\bf e}_1\times {\bf e}_2$ is perpendicular to the orbital plane. Since $A\neq B$, the satellite experiences a torque. Specifically, when points 1 and 2 are considered, the torque is
\begin{equation}
{\bf D}^{(1,2)}=\frac{d_{12}{\bf e}_1}{2}\times\left({\bf F}_1-{\bf F}_2\right),
\label{eqB1}
\end{equation}
where ${\bf F}_i=-GMm{\bf r}_ir_i^{-3}$ is the gravitational force acting on mass $i$. Using the law of cosines to the triangles $SHi$ ($i=1,2$) and the fact that $d_{12}\ll r$, one obtains
\begin{equation}
\frac{1}{r_i^3}\approx\frac{1}{r^3}\left(1\mp\frac{3d_{12}}{2r}\cos\alpha\right),
\label{eqB2}
\end{equation}
where the positive sign holds for $r_1$, while the negative for $r_2$. Inserting this into Eq.~(\ref{eqB1}), after some algebraic manipulations one arrives at
\begin{equation}
{\bf D}^{(1,2)}=\frac{6\pi^2 B\left(\frac{a}{r}\right)^3\sin2\alpha}{T^2}{\bf e}_3.
\label{eqB3}
\end{equation}
Similarly, the torque ${\bf D}^{(3,4)}$ can be calculated as
\begin{equation}
{\bf D}^{(3,4)}=-\frac{6\pi^2 A\left(\frac{a}{r}\right)^3\sin2\alpha}{T^2}{\bf e}_3.
\label{eqB4}
\end{equation}
The total torque is therefore ${\bf D}={\bf D}^{(1,2)}+{\bf D}^{(3,4)}$, and with Euler's second law: ${\bf D}=\dot{\bf L}=C\ddot{\theta}{\bf e}_3$ the equation of angular motion takes the form from Eq.~(\ref{eq11}).

\subsection{Introducing the second satellite}

The torques coming from an additional satellite and acting on the oblate moon are given by
\begin{equation}
{\bf D}^{(1,2)}_T=\frac{3Gm_1B\sin 2\alpha_1}{2r_{TH}^3}{\bf e}_3
\label{eqB5}
\end{equation}
and
\begin{equation}
{\bf D}^{(3,4)}_T=-\frac{3Gm_1A\sin 2\alpha_1}{2r_{TH}^3}{\bf e}_3,
\label{eqB6}
\end{equation}
where $m_1$ is the mass of the second satellite. The total torque acting on an oblate body is then
\begin{equation}
\begin{array}{l}
{\bf D}_{\rm tot}={\bf D}+{\bf D}_T={\bf D}^{(1,2)}+{\bf D}^{(3,4)}+{\bf D}^{(1,2)}_T+{\bf D}^{(3,4)}_T \\
\textcolor{white}{{\bf D}_{\rm tot}} = \frac{3(B-A)}{2}\left(\frac{GM\sin 2\alpha}{r^3}+\frac{Gm_1\sin 2\alpha_1}{r_{TH}^3}\right){\bf e}_3,
\end{array}
\label{eqB7}
\end{equation}
which after inserting the formula for $\alpha_1$ from Eq.~(\ref{eq16}), and noting that $Gm_1=GM\frac{m_1}{M}=n^2a^3\frac{m_1}{M}$, leads via Euler's second law to the equation of motion from Eq.~(\ref{eq21}).


\begin{thebibliography}{}
\bibitem[\protect\citeauthoryear{Alligood et al.}{2000}]{alligood} Alligood, K. T., Sauer, T. D., \& Yorke, J. A.: Chaos. An Introduction to Dynamical Systems. Springer, Berlin (2000)

\bibitem[\protect\citeauthoryear{Ascher \& Petzold}{1998}]{ascher} Ascher, U. M., \& Petzold, L. R.: Computer Methods for Ordinary Differential Equations and Differential-Algebraic Equations. SIAM, Philadelphia (1998)

\bibitem[\protect\citeauthoryear{Baker \& Gollub}{1996}]{baker} Baker, G., \& Gollub, J.: Chaotic Dynamics: An Introduction. Cambridge University Press (1996)

\bibitem[\protect\citeauthoryear{Beletskii}{1963}]{beletskii3} Beletskii, V. V.: The Motion of an Artificial Satellite about Its Center of Mass. New York: Plenum (1963)

\bibitem[\protect\citeauthoryear{Beletskii \& Levin}{1981}]{beletskii2} Beletskii, V. V., \& Levin, E. M.: Correctness of averaging in the plane problem of the resonance of Venus. Sov. Astron. \textbf{25}, 234 (1981)

\bibitem[\protect\citeauthoryear{Beletskii, Pivovarov \& Starostin}{1996}]{beletskii} Beletskii, V. V., Pivovarov, M. L., \& Starostin, E. L.: Regular and chaotic motions in applied dynamics of a rigid body. Chaos \textbf{6}, 155 (1996)

\bibitem[\protect\citeauthoryear{Bevilacqua et al.}{1980}]{bevilacqua} Bevilacqua, R., Menchi, O., Milani, A., Nobili, A. M., Farinella, P.: Resonances and close approaches. I - The Titan-Hyperion case. Moon and the Planets \textbf{22}, 141 (1980)

\bibitem[\protect\citeauthoryear{Black}{1995}]{black} Black, G. J., Nicholson, P. D., \& Thomas, P. C.: Hyperion: Rotational dynamics. Icarus \textbf{117}, 149 (1995)

\bibitem[\protect\citeauthoryear{Bond}{1848}]{bond} Bond, W. C.: Discovery of a new satellite of Saturn. Mon. Not. R. Astron. Soc. \textbf{9}, 1 (1848)

\bibitem[\protect\citeauthoryear{Boyd et al.}{1994}]{boyd} Boyd, P. T., Mindlin, G. B., Gilmore, R., \& Solari H. G.: Topological analysis of chaotic orbits: Revisiting Hyperion. Astrophys. J. \textbf{431}, 425 (1994)

\bibitem[\protect\citeauthoryear{Celletti \& Chierchia}{1998}]{celletti1} Celletti, A., \& Chierchia, L.: KAM stability estimates in Celestial Mechanics. Planet. Space Sci. \textbf{46}, 1433 (1998)

\bibitem[\protect\citeauthoryear{Celletti \& Chierchia}{2000}]{celletti2} Celletti, A., \& Chierchia, L.: Hamiltonian Stability of Spin Orbit Resonances in Celestial Mechanics. Celest. Mech. Dyn. Astr. \textbf{76}, 229 (2000)

\bibitem[\protect\citeauthoryear{Celletti \& Chierchia}{2008}]{celletti3} Celletti, A., \& Chierchia, L.: Measures of basins of attraction in spin-orbit dynamics. Celest. Mech. Dyn. Astr. \textbf{101}, 159 (2008)

\bibitem[\protect\citeauthoryear{Chirikov}{1979}]{chirikov} Chirikov, B. V.: A universal instability of many-dimensional oscillator systems. Phys. Rep. \textbf{52}, 263 (1979)

\bibitem[\protect\citeauthoryear{Colombo, Franklin \& Shapiro}{1974}]{colombo} Colombo, G., Franklin, F. A., Shapiro, I. I.: On the formation of the orbit-orbit resonance of Titan and Hyperion. Astron. J. \textbf{79}, 61 (1974)

\bibitem[\protect\citeauthoryear{Correia et al.}{2015}]{correia} Correia, A. C. M., Leleu, A., Rambaux, N., \& Robutel, P.: Spin-orbit coupling and chaotic rotation for circumbinary bodies. Application to the small satellites of the Pluto-Charon system. Astron. Astrophys. \textbf{580}, L14 (2015)

\bibitem[\protect\citeauthoryear{Crawford}{1991}]{crawford} Crawford, J. D.: Introduction to bifurcation theory. Rev. Mod. Phys. \textbf{63}, 991 (1991)

\bibitem[\protect\citeauthoryear{Danby}{1962}]{danby} Danby, J. M. A.: Fundamentals of Celestial Mechanics. The MacMillan Company, New York (1962)

\bibitem[\protect\citeauthoryear{Devyatkin et al.}{2002}]{devyatkin} Devyatkin, A. V., Gorshanov, D. L., Gritsuk, A. N., Mel'nikov, A. V., Sidorov, M. Yu., \& Shevchenko, I. I.: Observations and Theoretical Analysis of Lightcurves of Natural Satellites of Planets. Solar Syst. Res. \textbf{36}, 248 (2002)

\bibitem[\protect\citeauthoryear{Dourneau}{1993}]{dourneau} Dourneau, G.: Orbital elements of the eight major satellites of Saturn determined from a fit of their theories of motion to observations from 1886 to 1985. Astron. Astrophys. \textbf{267}, 299 (1993)

\bibitem[\protect\citeauthoryear{Feigenbaum}{1979}]{feigenbaum} Feigenbaum, M. J.: The universal metric properties of nonlinear transformations. J. Statist. Phys. \textbf{21}, 669 (1979)

\bibitem[\protect\citeauthoryear{Goldreich \& Peale}{1966}]{goldreich} Goldreich, P., \& Peale, S.: Spin-orbit coupling in the solar system. Astron. J. \textbf{71}, 425 (1966)

\bibitem[\protect\citeauthoryear{Grassberger \& Procaccia}{1983}]{grassberger} Grassberger, P., \& Procaccia, I.: Measuring the strangeness of strange attractors. Physica D \textbf{9}, 189 (1983)

\bibitem[\protect\citeauthoryear{Grassberger}{1986}]{grassberger2} Grassberger, P.: Do climatic attractors exist? Nature \textbf{323}, 609 (1986)

\bibitem[\protect\citeauthoryear{Greiner}{2010}]{greiner} Greiner, W.: Classical Mechanics. Systems of Particles and Hamiltonian Dynamics. Springer, Berlin Heidelberg (2010)

\bibitem[\protect\citeauthoryear{Harbison, Thomas \& Nicholson}{2011}]{harbison} Harbison, R. A., Thomas, P. C., \& Nicholson, P. C.: Rotational modeling of Hyperion. Celest. Mech. Dyn. Astr. \textbf{110}, 1 (2011)

\bibitem[\protect\citeauthoryear{Hausdorff}{1919}]{hausdorff} Hausdorff, F.: Dimension und \"au{\ss}eres Ma{\ss}. Mathematische Annalen \textbf{79}, 157 (1919)

\bibitem[\protect\citeauthoryear{Hicks, Buratti \& Basilier}{2008}]{hicks} Hicks, M. D., Buratti, B. J., \& Basilier, E. N.: BVR photometry of Hyperion near the time of the 2005 Cassini encounter. Icarus \textbf{193}, 352 (2008)

\bibitem[\protect\citeauthoryear{Jacobson et al.}{2006}]{jacobson} Jacobson, R. A., Antreasian, P. G., Bordi, J. J., Criddle, K. E., Ionasescu, R., Jones, J. B., Mackenzie, R. A., Meek, M. C., Parcher, D., Pelletier, F. J., Owen, W. M., Jr., Roth, D. C., Roundhill, I. M., \& Stauch, J. R.: The Gravity Field of the Saturnian System from Satellite Observations and Spacecraft Tracking Data. Astron. J. \textbf{132}, 2520 (2006)

\bibitem[\protect\citeauthoryear{Khan, Sharma \& Saha}{1998}]{khan} Khan, A., Sharma, R., \& Saha, L. M.: Chaotic Motion of an Ellipsoidal Satellite. I. Astrophys. J. \textbf{116}, 2058 (1998)

\bibitem[\protect\citeauthoryear{Klavetter}{1989a}]{klav} Klavetter, J. J.: Rotation of Hyperion. I - Observations. Astron. J. \textbf{97}, 570 (1989a)

\bibitem[\protect\citeauthoryear{Klavetter}{1989b}]{klav2} Klavetter, J. J.: Rotation of Hyperion. II - Dynamics. Astron. J. \textbf{98}, 1855 (1989b)

\bibitem[\protect\citeauthoryear{Kouprianov \& Shevchenko}{2003}]{kouprianov1} Kouprianov, V. V., \& Shevchenko, I. I.: On the chaotic rotation of planetary satellites: The Lyapunov exponents and the energy. Astron. Astrophys. \textbf{410}, 749 (2003)

\bibitem[\protect\citeauthoryear{Kouprianov \& Shevchenko}{2005}]{kouprianov2} Kouprianov, V. V., \& Shevchenko, I. I.: Rotational dynamics of planetary satellites: A survey of regular and chaotic behavior. Icarus \textbf{176}, 224 (2005)

\bibitem[\protect\citeauthoryear{Kuznetsov}{2004}]{kuznetsov} Kuznetsov, Y.: Elements of Applied Bifurcation Theory. Springer, New York (2004)

\bibitem[\protect\citeauthoryear{Lassel}{1848}]{lassel} Lassel, W.: Discovery of a new Satellite of Saturn. Mon. Not. R. Astron. Soc. \textbf{8}, 195 (1848)

\bibitem[\protect\citeauthoryear{Lichtenberg \& Lieberman}{1992}]{lichtenberg} Lichtenberg, A. J., \& Lieberman, M. A.: Regular and Chaotic Dynamics. Springer, New York (1992)

\bibitem[\protect\citeauthoryear{Maciejewski}{1995}]{maciejewski} Maciejewski, A. J.: Non-Integrability of the Planar Oscillations of a Satellite. Acta Astron. \textbf{45}, 327 (1995)

\bibitem[\protect\citeauthoryear{Mandelbrot}{1983}]{mandelbrot} Mandelbrot, B.: The Fractal Geometry of Nature. W. H. Freeman and Company, New York (1983)

\bibitem[\protect\citeauthoryear{May}{1976}]{may} May, R. M.: Simple mathematical models with very complicated dynamics. Nature \textbf{261}, 459 (1976)

\bibitem[\protect\citeauthoryear{Melnikov \& Shevchenko}{2010}]{melnikov} Melnikov, A. V., \& Shevchenko V. V.: The rotation states predominant among the planetary satellites. Icarus \textbf{209}, 786 (2010)

\bibitem[\protect\citeauthoryear{Melnikov}{2014}]{melnikov14} Melnikov, A. V.: Conditions for appearance of strange attractors in rotational dynamics of small planetary satellites. Cosm. Res. \textbf{52}, 461 (2014)

\bibitem[\protect\citeauthoryear{Ott}{2002}]{ott} Ott, E.: Chaos in dynamical systems. Cambridge University Press, Cambridge (2002)

\bibitem[\protect\citeauthoryear{Peale}{1976}]{peale} Peale, S. J.: Orbital resonances in the solar system. Annu. Rev. Astron. Astr. \textbf{14}, 215 (1976)

\bibitem[\protect\citeauthoryear{Peitgen, J\"urgens \& Saupe}{2004}]{peitgen} Peitgen, H.-O., J\"urgens, H., \& Saupe, D.: Chaos and Fractals. New Frontiers of Science. Springer, New York (2004)

\bibitem[\protect\citeauthoryear{Pomeau \& Manneville}{1980}]{pomeau} Pomeau, Y., \& Manneville, P.: Intermittent transition to turbulence in dissipative dynamical systems. Commun. Math. Phys. \textbf{53}, 643 (1980)

\bibitem[\protect\citeauthoryear{Rein et al.}{2012}]{rein} Rein, H., Payne, M. J., Veras, D., \& Ford, E. B.: Traditional formation scenarios fail to explain 4:3 mean motion resonances. Mon. Not. R. Astron. Soc. \textbf{426}, 187 (2012)

\bibitem[\protect\citeauthoryear{Seidelmann et al.}{2007}]{seidelmann} Seidelmann, P. K., Archinal, B. A., A'Hearn, M. F., Conrad, A., Consolmagno, G. J., Hestroffer, D., Hilton, J. L., Krasinsky, G. A., Neumann, G., Oberst, J., Stooke, P., Tedesco, E. F., Tholen, D. J., Thomas, P. C., \& Williams, I. P.: Report of the IAU/IAG Working Group on cartographic coordinates and rotational elements: 2006. Celest. Mech. Dyn. Astr. \textbf{98}, 155 (2007)

\bibitem[\protect\citeauthoryear{Shevchenko}{2002}]{shevchenko1} Shevchenko, I. I.: Maximum Lyapunov Exponents for Chaotic Rotation of Natural Planetary Satellites. Cosmic Res. \textbf{40}, 296 (2002)

\bibitem[\protect\citeauthoryear{Shevchenko \& Kouprianov}{2002}]{shevchenko} Shevchenko, I. I., \& Kouprianov V. V.: On the chaotic rotation of planetary satellites: The Lyapunov spectra and the maximum Lyapunov exponents. Astron. Astrophys. \textbf{394}, 663 (2002)

\bibitem[\protect\citeauthoryear{Smith et al.}{1982}]{smith} Smith, B. A., Soderblom, L., Batson, R. M., Bridges, P. M., Inge, J. L., Masursky, H., Shoemaker, E., Beebe, R. F., Boyce, J., Briggs, G., Bunker, A., Collins, S. A., Hansen, C., Johnson, T. V., Mitchell, J. L., Terrile, R. J., Cook, A. F., Cuzzi, J. N., Pollack, J. B., Danielson, G. E., Ingersoll, A. P., Davies, M. E., Hunt, G. E., Morrison, D., Owen, T., Sagan, C., Veverka, J., Strom, R., Suomi, V. E.: A new look at the Saturn system - The Voyager 2 images. Science \textbf{215}, 504 (1982)

\bibitem[\protect\citeauthoryear{Spohn, Johnson \& Breuer}{2014}]{ency} Spohn, T., Johnson T., \& Breuer, D.: Encyclopedia of the Solar System (Third Edition). Elsevier, Boston (2014)

\bibitem[\protect\citeauthoryear{Stellmacher}{1999}]{stellmacher} Stellmacher, I.: Periodic solutions for resonance 4/3: Application to the construction of an intermediary orbit for Hyperion's motion. Celest. Mech. Dyn. Astr. \textbf{75}, 185 (1999)

\bibitem[\protect\citeauthoryear{Strugnell \& Taylor}{1990}]{strugnell} Strugnell, P. R., \& Taylor, D. B.: A catalogue of ground-based observations of the eight major satellites of Saturn, 1874-1989. Astrophys. Space Sci. \textbf{83}, 289 (1990)

\bibitem[\protect\citeauthoryear{Tabor}{1989}]{tabor} Tabor, M.: Chaos and Integrability in Nonlinear Dynamics: An Introduction. Wiley, New York (1989)

\bibitem[\protect\citeauthoryear{Thomas et al.}{2007}]{thomas07} Thomas, P. C., Armstrong, J. W., Asmar, S. W., Burns, J. A., Denk, T., Giese, B., Helfenstein, P., Iess, L., Johnson, T. V., McEwen, A., Nicolaisen, L., Porco, C., Rappaport, N., Richardson, J., Somenzi, L., Tortora, P., Turtle, E. P., \& Veverka, J.: Hyperion's sponge-like appearance. Nature \textbf{448}, 50 (2007)

\bibitem[\protect\citeauthoryear{Thomas}{2010}]{thomas10} Thomas, P. C.: Sizes, shapes, and derived properties of the saturnian satellites after the Cassini nominal mission. Icarus \textbf{208}, 395 (2010)

\bibitem[\protect\citeauthoryear{Tarnopolski}{2014}]{tarnopolski1} Tarnopolski, M.: On the fractal dimension of the Duffing attractor. Rom. Rep. Phys. \textbf{66}, 907 (2014)

\bibitem[\protect\citeauthoryear{Tarnopolski}{2015a}]{tarnopolski} Tarnopolski, M.: Nonlinear time-series analysis of Hyperion's lightcurves. Astrophys. Space Sci. \textbf{357}:160 (2015a)

\bibitem[\protect\citeauthoryear{Tarnopolski}{2015b}]{tarnopolski2} Tarnopolski, M.: Testing the anisotropy in the angular distribution of Fermi/GBM gamma-ray bursts. Preprint (\href{http://arxiv.org/abs/1512.02865}{arXiv:1512.02865}) (2015b)

\bibitem[\protect\citeauthoryear{Taylor}{1987}]{taylor} Taylor, D. B., Sinclair, A. T., \& Message, P. J.: Corrections to the theory of the orbit of Saturn's satellite Hyperion. Astron. Astrophys. \textbf{181}, 383 (1987)

\bibitem[\protect\citeauthoryear{Taylor}{1992}]{taylor2} Taylor, D. B.: A synthetic theory for the perturbations of Titan on Hyperion. Astron. Astrophys. \textbf{265}, 825 (1992)

\bibitem[\protect\citeauthoryear{Theiler}{1990}]{theiler} Theiler, J.: Estimating fractal dimension. J. Soc. Opt. Am. A \textbf{7}, 1055 (1990)

\bibitem[\protect\citeauthoryear{Wisdom, Peale \& Mignard}{1984}]{wisdom} Wisdom, J., Peale, S. J., \& Mignard, S.: The chaotic rotation of Hyperion. Icarus \textbf{58}, 137 (1984)

\end{thebibliography}
\end{document}